\documentclass[12pt,amsfonts]{article}
\usepackage{amsfonts}
\def\ts{\textstyle}
\def\t{\textstyle}        
\def\one{1\hskip-.37em 1}                

\def\half{{\textstyle{\frac{1}{2}}}}
\def\wbet{\widetilde{\beta}}
\def\wp{\widetilde p}
\def\wq{\widetilde q}
\def\wG{\widetilde G}

\def\quarter{\textstyle{\frac{1}{4}}}

\def\H{{\cal H}}

\def\p{\phi}

\def\H{{\cal H}}

\def\v{\vskip.3em}

\def\l{\lambda}

\def\pa{\overrightarrow{p}}
\def\Pa{\overrightarrow{P}}
\def\qa{\overrightarrow{q}}
\def\Qa{\overrightarrow{Q}}
\def\Ra{\overrightarrow{R}}
\def\Sa{\overrightarrow{S}}

\def\ra{\rightarrow}

\def\tint{{\textstyle\int}}

\def\s{\hskip.08em}
\def\d{\partial}

\def\b{\begin{eqnarray*}}  
\def\e{\end{eqnarray*}}    
\def\bn{\begin{eqnarray}}  
\def\en{\end{eqnarray}}   
\def\<{\langle}
\def\>{\rangle}

\def\no{\nonumber}

\def\k{\kappa}

\def\{{\lbrace}
\def\}{\rbrace}

\title{Completing Canonical Quantization, \\and Its Role in \\Nontrivial Scalar Field Quantization\footnote{Based on two separate lectures presented at the conference ``Stochastic and Infinite Dimensional Analysis'', Bielefeld, Germany, June 2013.}}
\author{John R. Klauder\footnote{Email: klauder@phys.ufl.edu}\\
Department of Physics and\\Department of Mathematics\\
University of Florida\\
Gainesville, FL 32611-8440}
\date{ }

\begin{document}
\maketitle
\begin{abstract}
The process of canonical quantization is redefined so that the classical and quantum theories coexist
when $\hbar>0$, just as they do in the real world. This analysis not only supports conventional procedures, it also reveals new quantization procedures that, among several examples, permit
nontrivial quantization of scalar field models such as $\phi^4_n$ for every spacetime dimension $n\ge2$.
\end{abstract}
\section{Conventional \& Enhanced Quantization}
\subsection{Conventional canonical quantization}
The standard recipe for canonical quantization is simply stated. For a single degree of freedom, one version
reads:\v

 {\bf Classical Theory:} Choose canonically conjugate, classical phase space coordinates $p$ and $q$, along with  a classical Hamiltonian $H_c(p,q)$, and adopt dynamical equations, i.e., Hamilton's equations of motion, that follow from the stationary variation of a classical (C) action functional given by
   \bn A_C\equiv \tint_0^T\s[\s p(t)\s{\dot q}(t)-H_c(p(t),q(t))\s]\,dt\;. \en\v

  {\bf Quantum Theory:} Promote the phase space coordinates to Hermitian operators, $p\ra P$ and $q\ra Q$, that satisfy Heisenberg's commutation relation $[\s Q\s,\s P\s]=i\hbar\one$. Choose a Hermitian Hamiltonian operator $\H(P,Q)$, with dynamical equations, i.e., Schr\"odinger's equation and its adjoint, that follow from the stationary variation of a quantum (Q) action functional given by
    \bn A_Q\equiv \tint_0^T\,\<\psi(t)|\s[\s i\hbar \s(\d/\d t)-\H(P,Q)\s]\s|\psi(t)\>\,dt\;, \en
    where $|\psi\>$---and its adjoint $\<\psi|\s$---denote vectors in a complex Hilbert space. \v

     {\bf Comments:} Clearly, the classical and quantum theories have several fundamental differences: For a single degree of freedom, classical phase space is two dimensional, while in quantum theory the Hilbert space is infinite dimensional. In the classical theory, one chooses $\hbar=0$, while in the quantum  theory $\hbar\simeq 10^{-27}erg\cdot sec$ as determined by experimental measurement. To relate the classical and quantum models, it is traditional to choose $\H(P,Q)=H_c(P,Q)$, modulo possible ${\cal O}(\hbar)$ corrections. While covariance under canonical coordinate transformations leads to many possible choices of coordinates, quantization results are generally better \cite{dirac} if the original coordinates $p$ and $q$ are chosen as ``Cartesian coordinates''---despite the fact that classical phase space is {\it not} endowed with a metric structure that would permit the identification of such coordinates.\v\v
    Although canonical quantization as sketched above is highly successful, there are certain cases where the so-defined quantum theory is less than satisfactory. We claim that the triviality of $\phi^4$ scalar field models in high enough spacetime dimensions are such cases. We aim to overcome that triviality.
\subsection{Enhanced canonical quantization}
    In the real world $\hbar>0$, and so there must be a formulation of the classical theory that accepts that fact \cite{enh}. The action functional for the quantum theory assumes that general variations of the Hilbert space vectors are possible. But suppose that is not the case, and we are able to  vary only certain Hilbert space vectors that can be varied {\it without} disturbing the system. One such variation involves translating the system to a new position, but according to Galilean covariance we can move the observer a corresponding amount instead of moving the system. Likewise we can imagine putting the system into uniform motion with a constant velocity, but again Galilean covariance asserts we can get the same result by putting the observer into uniform motion instead of the system. Thus if we assume that some normalized reference state $|\eta\>$ is relevant for our problem, e.g.,  under appropriate conditions, the ground state for our system, then we can imagine that we can vary the set of states denoted by
       \bn    |p,q\>\equiv e^{\t-iqP/\hbar}\,e^{\t ip\s Q/\hbar}\s|\eta\>\;, \en
       a set of states recognized as {\it canonical coherent states} \cite{skag-kla}, which offer translation of $|\eta\>$ to a new position by $q$ as well as to a new velocity as represented by a new momentum $p$. Here the operators $P$ and $Q$, which (for the present) we assume to be irreducible, obey the usual commutation relation $[\s Q,P\s]=i\hbar\one$, and, moreover, are chosen as {\it self adjoint}---a stronger condition than simply being Hermitian---which is necessary to generate unitary transformations that preserve the normalization of $|\eta\>$. It is convenient to choose $|\eta\>$ such that $\<\eta|\s P\s|\eta\>=\<\eta|\s Q\s|\eta\>=0$, called ``physical centering'', in which case $\<p,q|\s P\s|p,q\>=p$ and $\<p,q|\s Q\s|p,q\>=q$, two equations that determine the physical meaning of $p$ and $q$. Armed with this set of states, we declare as a classical observer that we can only vary $p$ and $q$ in the limited set of states $\{|p,q\>\}$, and thus
       we are led to a restricted (R) quantum action functional given by
          \bn A_{Q(R)}\hskip-1.3em&&\equiv \tint_0^T\,\<p(t),q(t)|\s[\s i\s\hbar\s(\d/\d t)-\H(P,Q)\s]\s|p(t),q(t)\>\,dt\no\\
               &&=\tint_0^T[\s p(t)\s{\dot q}(t)-H(p(t),q(t))\s]\,dt\;, \en
       where $\<p,q|\s i\hbar\s(\d/\d t)\s|p,q\>=\<\eta|\s[\s(P+p\s\one)\s{\dot q}-Q\s{\dot p}\s\s]\s|\eta\>=p\s\s{\dot q}$.
       Here we have introduced the important relation that
          \bn H(p,q)\hskip-1.3em&&\equiv \<p,q|\s \H(P,Q)\s|p,q\>\no\\
          &&=\<\eta|\s\H(P+p\one,Q+q\one)\s|\eta\>\no\\
          &&=\H(p,q)+{\cal O}(\hbar;p,q)\;.  \label{e1}\en
          Observe that the restricted quantum action functional $A_{Q(R)}$ resembles the classical action
          functional and differs from it only by the fact that $\hbar>0$. The conventional classical Hamiltonian may be obtained as
            \bn H_c(p,q)\equiv \lim_{\hbar\ra0}\, H(p,q)\;, \en
            but that limit may change the character of $H(p,q)$ in unphysical ways. The additional term ${\cal O}(\hbar;p,q)$ in (\ref{e1}) depends on the choice of $|\eta\>$, and generally would change if $|\eta\>$ is changed. This is to be expected since our limited set of states $\{|p,q\>\}$ involves  {\it projections} in Hilbert space, and different choices of $|\eta\>$ lead to different projections.

          Classical mechanics also involves canonical coordinate transformations which relate new canonical coordinates $(\wp,\wq)$ to our present coordinates $(p,q)$ by means of the one form
            \bn p\s\s dq=\wp\s\s d\wq+d\wG(\wp,\wq)\;.\en
         We define the coherent states to transform as scalars under such coordinate transformations such that $|p,q\>=|p(\wp,\wq),q(\wp,\wq)\>\equiv|\wp,\wq\>$, so that the restricted quantum action functional becomes
            \bn I_{Q(R})\hskip-1.3em&&=\tint_0^T\<\wp(t),\wq(t)|\s [\s i\hbar \s(\d/\d t)-\H(P,Q)\s]\s|\wp(t),\wq(t)\>\,dt\no\\
                &&=\tint_0^T\s[\s \wp(t)\s{\dot{\wq}}(t)+{\dot{\wG}}(\wp(t),\wq(t))-{\widetilde H}(\wp(t),\wq(t))\s]\,dt\;, \en
                where ${\widetilde H}(\wp,\wq)\equiv H(p,q)$, and which leads to the proper enhanced classical equations of motion without changing the quantum operators in any way.

         Next, let us return to the original coordinates $(p,q)$ and observe that in these canonical coordinates
          the relation (\ref{e1}) has {\it exactly} the feature characterized by the choice of $p$ and $q$ as ``Cartesian coordinates''. And indeed, we can now show that these are Cartesian coordinates after all. The Hilbert space norm generates a metric for vectors determined by $d(|\psi\>,|\phi\>)^2=\|\s|\psi\>-|\phi\>\s\|^2$. However, in quantum theory the overall phase of a vector carries no physics and we can instead consider the ray (R) metric determined by  $d_R(|\psi\>,|\phi\>)^2=\min_\alpha\|\s|\psi\>-e^{i\s\alpha}\s|\phi\>\s\|^2$. If we evaluate a rescaled version of the ray metric for two coherent states that are infinitesimally close to each other, we obtain (a.k.a., the Fubini-Study metric)
              \bn d\sigma_R(p,q)^2\hskip-1.3em&&\equiv (2\s\hbar)\s[\s \|\s d|p,q\>\s\|^2-|\s\<p,q|\s d|p,q\>\s|^2\s] \no\\
                 &&= (2/\hbar)\,[\s\<Q^2\>\s dp^2+\<\s P\s Q+Q\s P\s\> dp\s\s dq+\<\s P^2\s\>\s dq^2\s\s]\;, \label{FS} \en
                 where here $\<(\cdot)\>\equiv\<\eta|(\cdot)|\eta\>$. For a general choice of $|\eta\>$, the
                 two-dimensional space $\{p,q\}$ is always flat, and up to a linear coordinate transformation, this metric involves Cartesian coordinates. Specifically, for the common choice where $(\omega\s Q+iP)\s|\eta\>=0$, i.e., an oscillator ground state,  then
                   \bn d\sigma_R(p,q)^2=\omega^{-1}\s dp^2+\omega\s dq^2\;, \en
                     and $p$ and $q$ are indeed Cartesian coordinates according to the Hilbert space metric. Of course, we can now assign that metric to the classical phase space if so desired. It is in this sense that the coordinates $p$ and $q$ are Cartesian.

           {\bf Conventional canonical quantization is confirmed!} {\it It is important to appreciate that what has been shown so far is equivalent to the standard canonical quantization procedure!} Specifically, we have identified phase-space coordinates $p$ and $q$ that are indeed Cartesian coordinates and the quantum Hamiltonian operator is indeed the same function of the variables as is the classical Hamiltonian, modulo terms of order $\hbar$.

           However, there is more to the story.
\subsubsection*{Reducible canonical operators}
           Let us consider the example with a classical Hamiltonian given by
             \bn H_c( \pa,\qa)=\half\s(\s\pa^2+m_0^2\s\qa^2\s)+\l\s(\qa^2)^2 \;,\en
             where $\pa=\{p_1,p_2,\ldots,p_N\}$, and  $\qa=\{q_1,q_2,\ldots,q_N\}$, with $N\le\infty$.
             Here, $\pa^2\equiv\Sigma_{n=1}^N\s p_n^2$, $\,\qa^2\equiv\Sigma_{n=1}^N\s q_n^2$, and for $N=\infty$ we require that $\pa^2+\qa^2<\infty$. It is clear that this model is invariant under orthogonal rotations $\pa\ra O\pa$, $\qa\ra O\qa$, where $O\in {\bf O}(N,{\mathbb{R}})$,
             and such models are called {\it Rotationally Symmetric} models \cite{rs}.
             As a consequence of rotational invariance, every solution is equivalent to a solution for $N=1$ if $\pa\s\|\qa$
             at time $t=0$, or to a solution for $N=2$ if $\pa\!\not\!\|\s \qa$ at time $t=0$. Moreover, solutions for $N=\infty$ may be derived from those for $N<\infty$ by the limit $N\ra\infty$, provided we maintain
             $\pa^2+\qa^2<\infty$.

            A conventional canonical quantization begins with $\pa\ra\Pa$, $\qa\ra\Qa$, which are irreducible operators that obey  $[Q_l,P_n]=i\hbar\delta_{l.n}\one$ as the only non-vanishing commutation relation. For a free model, with mass $m$ and $\l=0$, the quantum Hamiltonian
            $\H_0=\half
            :( \Pa^2+m^2\s\Qa^2):$, where $:(\cdot):$ denotes normal ordering, has the feature that the Hamiltonian operator for $N=\infty$ is obtained as the limit of those for which $N<\infty$.
            Moreover, with the ground state $|0\>$ of the Hamiltonian operator chosen as the fiducial vector for canonical coherent states, namely,  \bn |\pa,\qa\>=\exp[-i\qa\cdot\Pa/\hbar]\s\exp[i\pa\cdot\Qa/\hbar]\s|0\>\;,\en
            it follows that
              \bn \<\pa,\qa|\s\half:(\Pa^2+m^2\s\Qa^2):|\pa,\qa\>=\half(\pa^2+m^2\s\qa^2)=H_0\pa,\qa)\en
              as desired, for all $N\le\infty$.

              However,  canonical quantization of the interacting models with $\l>0$  leads to trivial results for $N=\infty$. To show this we assume that the Schr\"odinger representation of the ground state of an interacting model is real, unique, and rotationally invariant. As a consequence, the characteristic function (i.e., the Fourier transform) of the ground-state distribution has the form (note: $|f|^2\equiv\Sigma_{n=1}^N\s f_n^2$ and $r^2\equiv\Sigma_{n=1}^N\s x_n^2$)
                \bn C_N(\overrightarrow{f})\hskip-1.3em&&=\int e^{\t i\Sigma_{n=1}^N f_n\s x_n/\hbar}\,\Psi_0(r)^2\,\Pi_{n=1}^N\s dx_n\no\\
                  &&=\int e^{\t i|f|\s r\s\cos(\theta)/\hbar}\,\Psi_0(r)^2\,r^{N-1}\, dr\s\sin(\theta)^{N-2}\, d\theta\,d\Omega_{N-3}\no\\
                  &&\simeq M'\int e^{\t-|f|^2\s r^2/2(N-2)\hbar^2}\,\Psi_0(r)^2\,r^{N-1}\, dr\s d\Omega_{N-3}\no\\
                  &&\ra\int_0^\infty e^{\t-b\s |f|^2/\hbar}\,w(b)\,db \en
                  assuming convergence, where a steepest descent integral has been performed for $\theta$, and in the last line we have taken the limit $N\ra\infty$. Additionally,
                  $w(b)\ge0$, and $\int_0^\infty w(b)\s db=1$. This is the result based on symmetry. Uniqueness of the ground state then ensures that $w(b)=\delta(b-1/4{\widetilde m})$, for some $\widetilde{m}>0$, implying that the quantum theory is that of a free theory, i.e., {\it the quantum theory is trivial! In addition, the classical limit of the resultant quantum theory is a free theory, which differs from the original, nonlinear classical theory.}

                  The way around this unsatisfactory result is to let the representations of $\Pa$ and $\Qa$ be {\it reducible}. The weak correspondence principle, namely
                  $ H(\pa,\qa)\equiv \<\pa,\qa|\s\H\s|\pa,\qa\>$, ensures that the enhanced classical Hamiltonian depends only on the proper variables. A detailed study \cite{rs} of the proper reducible representation, still in accord with the argument above that limits the ground-state functional form to a Gaussian, leads to the following formulation. Let $\Ra$ and $\Sa$ represent a new set of operators, independent of the former operators, and which obey the commutation relation $[S_l,R_n]=\i\hbar\delta_{l,n}\one$. We introduce two Hamiltonian operators:
                    \bn &&\H_{0\,PQ}\equiv\half:(\Pa^2+m^2(\Qa+\zeta\Sa)^2\s):\;,\no\\
                        &&\H_{0\,RS}\equiv\half:(\Ra^2+m^2(\Sa+\zeta\Qa)^2\s):\;,\en
                        where $0<\zeta<1$.
                   These two operators have a common, unique, Gaussian ground state $|0,0;\zeta\>$.
                   Let new coherent states, which span the Hilbert space of interest, be defined with this ground state as the fiducial vector, as given by
                   \bn|\pa,\qa\>\equiv \exp[-i\qa\cdot\Pa/\hbar]\s\exp[i\pa\cdot\Qa/\hbar]\s|0,0;\zeta\>\;,\en
                   and it follows that
                   \bn &&\<\pa,\qa|\s \H_{0\,PQ}+\H_{0\,RS}+4v\s:\H_{0\,RS}^2:\s|\pa,\qa\>\no\\
                       &&\hskip2em=\half[\pa^2+m^2(1+\zeta^2)\s\qa^2]+v\s\zeta^4\s m^4\s (\qa^2)^2\no\\
                       &&\hskip2em\equiv \half(\pa^2+m_0^2\s \qa^2)+\l\s(\qa^2)^2\en
                       as required. This example shows that enhanced quantization techniques that make use of reducible kinematical operator representations can lead to
                       a nontrivial and fully satisfactory solution to certain problems.

                       The next section illustrates yet another procedure that serves to generalize canonical quantization.

\subsection{Enhanced affine quantization}
\subsubsection*{Affine variables and their algebra}
           We return to the study of a single degree of freedom.
          Importantly,  the canonical operators $P$ and $Q$, which have the whole real line for their spectrum and  satisfy the Heisenberg commutation rule $[\s Q\s,\s P\s]=i\hbar\one$, imply a second commutation relation as well. If we multiply the Heisenberg commutator by $Q$, we find $i \hbar \s Q=[Q,P]\s Q=[Q,P\s Q]$,  and finally the Lie algebra
            \bn [\s Q\s, \s D\s]=i\s\hbar\s Q\;,\hskip3em D\equiv\half\s( P\s Q+Q\s P)\;.  \label{e4} \en
            The variables $D$ and $Q$ are called {\it affine coordinates} and the commutation relation (\ref{e4}) is
            called an {\it affine commutation relation}. Clearly $D$ has the dimensions of $\hbar$, and we will find it convenient to choose $Q$ as dimensionless (or consider $Q/q_0$ and choose units so that $q_0=1$). If the representation for $P$ and $Q$ is {\it irreducible}, then the representation for $D$ and $Q$ is {\it reducible}. The irreducible sub-representations of $D$ and $Q$ are one where $Q>0$ and a similar, second one where $Q<0$; a third representation with $Q=0$ is less important. Initially, let us consider the irreducible representation where $Q>0$.

            If $Q>0$ is a self-adjoint operator, then it follows that $P$, although Hermitian, can never be self adjoint. Thus, $P$ can not serve as the generator of unitary transformations, so canonical coherent states do not exist in this case. However, although $P$ can not be self adjoint, the operator $D$ can be self adjoint; hence we choose a different algebra made from the affine variables. We choose a new normalized fiducial vector $|\eta\>$ and introduce a set of {\it affine coherent states} \cite{skag-kla}, which are defined by
              \bn |p,q\>\equiv e^{\t ip\s Q/\hbar}\,e^{\t-i\ln(q)\s D/\hbar}\s|\eta\>\;, \en
              for all $(p,q)\in\mathbb{R}\times\mathbb{R}^+$, i.e., $q>0$. While a self-adjoint $P$ serves to {\it translate} $Q$, e.g., $e^{iqP/\hbar}\s Q\s e^{-iqP/\hbar}=Q+q\one$, it follows that
          a self-adjoint $D$ serves to {\it dilate} $Q$, e.g., $e^{i\ln(q)D/\hbar}\s Q\s e^{-i\ln(q)D/\hbar}=q\s Q$; as already partially noted, it is useful to treat $q$ and $Q$ as dimensionless. If we choose $|\eta\>$ so that $[\wbet\s (Q-1)+i\s D\s]\s|\eta\>=0$---a rough analog of $(\omega\s Q+iP)\s|\eta\>=0$ for the Heisenberg algebra---it follows that $\<\eta|Q|\eta\>=1$ and $\<\eta|D|\eta\>=0$. Moreover, $\<p,q|Q|p,q\>=q$ as well as $\<p,q|D|p,q\>=p\s\s q$.

          It is also important to consider reducible affine operator representations as well. In this case,
          we introduce a fiducial vector $|\eta\>= |\eta_+\>\oplus |\eta_-\>$ and $Q=Q_+\oplus Q_-$,where $\<\eta_\pm|Q_\pm|\eta_\pm\>=\pm1$. We introduce reducible affine coherent states given  by
              \bn |p,q\s\pm \>\equiv |p,q_+\>\oplus |p,q_-\>\;,\en
              where $p\in\mathbb{R}$, $\pm\s\s q_\pm>0$, $\pm \s\s Q_\pm>0$, and
            \bn  |p,q_\pm\>\equiv e^{\t ip\s Q_\pm/\hbar}\,e^{\t-i\ln(|q|)\s D/\hbar}\s|\eta_\pm\> \;.\en
            In particular, with the indicated choice of the fiducial vector, the coherent state overlap function is given, for
          separate $\pm$ in each vector, by
    \bn &&\<p',q'\pm\s|p,q\s\pm \>\no\\
    &&\hskip2em=\theta(q'q)\,[\s\half\s(\sqrt{q'/q}+\sqrt{q/q'}\s\s)
   +i\s\half\s\sqrt{q'\s q}\s(p'-p)/\wbet\s]^{-2\wbet/\hbar}\;,\en
   where $\theta(y)\equiv1$ if $y>0$ and $\theta(y)\equiv0$ if $y<0$.

   \subsubsection*{Affine quantization as canonical quantization}
     Given a quantum action functional, once again we assume that we can only vary a subset of Hilbert space vectors, in particular, either the irreducible affine coherent states $\{|p,q\>\}$ or the reducible affine coherent states $\{|p,q\s\pm\>\}$. This leads to two versions of the restricted quantum action  functional:
         \bn A_{Q(R)}\hskip-1.3em&&=\tint_0^T\,\<p(t),q(t)|\s[\s i\hbar(\d/\d t)-\H'(D,Q)\s]\s|p(t),q(t)\>\,dt\no\\
            &&=\tint_o^T\s[\s - q(t)\s{\dot p}(t)-H(p(t),q(t))\s]\,dt\;,\en
            in which $q(t)>0$, as well as, for identical $\pm$ in both vectors,
            \bn A_{Q(R\pm)}\hskip-1.3em&&=\tint_0^T\,\<p(t),q(t)\pm|\s[\s i\hbar(\d/\d t)-\H'(D,Q)\s]\s|p(t),q(t)\s\pm\>\,dt\no\\
            &&=\tint_0^T\s[\s -q(t)\s{\dot p}(t)-H(p(t),q(t))\s]\,dt\;,\en
            where now $|q(t)|>0$. Note that this latter case is especially useful if the Hamiltonian has a singularity at  $q=0$. {\it However, the important point is:}  {\bf The restricted quantum action functional, based on affine coherent states, is again identical to the form of a canonical classical system, enhanced, because in these equations, $\hbar>0$!} {\bf Stated otherwise, enhanced affine quantization effectively serves as enhanced canonical quantization.}

            To complete the story we observe that
             \bn H(p,q)\hskip-1.3em&&\equiv\<p,q\pm|\H'(D,Q)\s|p,q\s\pm\>\no\\
             &&=\<\H'(D+p|q|\s Q,|q|\s Q)\>=\H'(p\s q,q)+{\cal O}(\hbar;p,q)\;,\en
             as well as
             \bn H(p,q)\hskip-1.3em&&\equiv\<p,q\pm|\s\H(P,Q)\s|p,q\s\pm\>\no\\
             &&=\<\s\H(P/|q|+p,|q|\s Q)\s\>=\H(p,q)+{\cal O}(\hbar;p,q)\;.\en

             It is important to appreciate that
             the coordinates $(p,q)$ used in the affine coherent states can be changed to a new set of  coordinates $(\wp,\wq)$ in the very same manner as was the case for the coordinates used in the canonical coherent states.
     While the phase-space geometry induced by the canonical coherent states (\ref{FS}) led to a flat space,
             the Fubini-Study metric for the affine coherent states, given by
                \bn d\sigma_R(p,q)^2= {\tilde\beta}^{-1}q^2\s dp^2+{\tilde\beta}\s q^{-2}\s dq^2\;,\en
                corresponds to a different phase-space geometry, namely,
                 a space of constant negative curvature, $-2/{\tilde\beta}$. However, this difference in induced phase-space geometry does not affect the suitability of enhanced affine quantization to serve as enhanced canonical quantization.

    \subsubsection*{A simple example}
             A simple example serves to illustrate the power of enhanced affine quantization. As the classical Hamiltonian, we choose
                $H_c(p,q)=p^2/2m - e^2/|q|$,
                which we call a one-dimensional `hydrogen atom' problem. Classical solutions fall into the singularity in a finite time. For the Hamiltonian operator we choose
                $\H= P^2/2m-e^2/|Q|$, and thus the enhanced affine classical Hamiltonian is given by
                   \bn H(p,q)= p^2/2m- C/|q|+ C'/q^2\;, \en
                   where $C\equiv\s e^2\<|Q|^{-1}\>\propto e^2$ and $C'\equiv\s\<P^2\>/2m\propto \hbar^2/m$. Observe that $C'\simeq \hbar^2/(m\s e^2)\s C$, {\it a ratio that is seen to be the Bohr radius!}. Thus we see that the enhanced affine classical Hamiltonian prevents all singularities and has a stable minimum at a distance from the singularity of approximately the Bohr radius. Since $\hbar>0$ in the real world, we are led to claim that the enhanced classical Hamiltonian is `more physical' than the conventional classical Hamiltonian with which we started, for which $\hbar=0$. It would be hard to achieve the same features from an enhanced canonical quantization.\v\v

                   It is important to emphasize that for enhanced affine quantization of Hamiltonians of the general form $P^2+V(Q)$, the enhanced classical Hamiltonian always contains a term of the form $q^{-2}$ with a coefficient proportional to $\hbar^2$. This fact anticipates the naturalness of an actual term proportional to $\hbar^2$ and involving inverse squared operators in the quantum Hamiltonian of other models,  suggesting that it
                   may not be out of place.

                   These musings provide an important clue for the next topic of discussion.
  \section{Scalar Field Quantization \\Without Divergences}
               We next consider a special class of infinitely-many degrees-of-freedom problems associated with a covariant scalar field. For $\phi^4_n$ models, standard canonical quantization procedures have obtained self-consistent solutions for spacetime dimensions $n=2,3$, e.g., \cite{GJ}, but those same methods have failed to provide suitable results for $n\ge4$, leading instead to trivial quantum solutions equivalent to (generalized) free theories.
               It is our believe that for higher spacetime dimensions ($n\ge4$), we can find nontrivial solutions by choosing affine quantization procedures, and that this technique also leads to new solutions for $n=2,3$ as well, which exhibit compatibility for `mixed models' in ways we will describe. The reason behind this believe is based on the form of the ground-state distribution we are led to, which has integrable singularities when certain field values are zero. As we shall see, this particular form of the ground-state distribution has the decided advantage that a perturbation series for the interaction does {\it not} exhibit divergences! Our ground-state wave function
               with square-integrable singularities when certain fields vanish leads to terms in the quantum Hamiltonian proportional to $\hbar^2$, and in our case involve inverse squared field operators.
                Although an affine quantization is in order, the insight outlined above for these models means that  we can proceed to develop the theory in a more direct manner \cite{JRK}.
   \subsection{Free  vs.~pseudofree models}
               It is self evident that $\lim_{g\ra0} (\s A_0+g\s A_I\s)=A_0$, except when it is false. Consider the action functional for an anharmonic oscillator given by
                 \bn A_g=\tint_0^T\,\{\s\half[\s {\dot y}(t)^2-y(t)^2]-g\s y(t)^{w}\s\}\,dt\;,\en
                 and the domain of functions allowed by this expression. If the exponent $w=+4$, then the limit as $g\ra0$ leads to the free action $A_0=\tint_0^T\s\half\s[{\dot y}(t)^2-y(t)^2\s]\s dt$, but if $w=-4$ that is {\it not} the case. Instead, when $w=-4$, the limit as $g\ra0$ is
                 $A'_0\equiv\tint_0^T\s\half\s[{\dot y}(t)^2-y(t)^2\s]\s dt$ {\it supplemented with the requirement that} $\tint_0^T\s y(t)^{-4}\,dt<\infty$, which is a fundamentally different domain from that of the free action functional. We refer to the theory described by $A'_0$ as a {\it pseudofree theory}.
                 A pseudofree theory is the one that is continuously connected to the interacting theories. The pseudofree theory may coincide with the usual free theory, as is the case when $w=+4$, but when $w=-4$, the pseudofree and the free theories are different. This distinction applies to the quantum theories as well. In particular, the free and pseudofree quantum theories are identical for
                $w=+4$ and distinct for $w=-4$. If one considered making a perturbation analysis of the interacting theory, one would have to start with the pseudofree theory and {\it not} with the free theory when these two theories differ.

                A rather similar situation applies to scalar fields. Consider the classical action given by
                \bn A_{g_0}=\tint (\s\half\s\{\s[{\d_\mu\phi}(t,x)]^2 -m_0^2\s\phi(t,x)^2\s\}-g_0\s\phi(t,x)^4\s)\,dt\s\s d^s\!x\;,  \en
                where $x\in\mathbb{R}^s$, $s\ge 1$. Provided $m_0>0$ and $g_0>0$, a multiplicative inequality \cite{lady,book1} implies, for $n=s+1$, and now where $x\in\mathbb{R}^n$,  that
                   \bn \{\s g_0\s\tint \s\phi(x)^4\,d^n\!x\}^{1/2}\le C''\s\tint\{\s[\s{\nabla\phi}(x)^2]+m_0^2\s\phi(x)^2\s\}\,d^n\!x\;,\label{w2}\en
                 where $C''=(4/3)[g_0^{1/2}\,m_0^{(n-4)/2}]$ if $n\le4$, while $C''=\infty$ if $n\ge5$, which in the latter case means that there are fields for which the left side of (\ref{w2}) diverges, but the right side is finite. [Remark: The divergent cases are {\it exactly} the {\it non}-renormalizable models when quantized---and that fact holds true for all the non-renormalizable models of the form $\p^p_n$ as well \cite{book1}!] Thus, when $n\ge5$, (\ref{w2}) ensures that the classical pseudofree model is different from the classical free model; and thus we expect that the quantum pseudofree and quantum free models are also different. We will be guided in choosing the pseudofree model by the requirement that it connects smoothly with the interacting models.
    \subsection{Choosing the pseudofree model}
                Formally, the action functional determines the quantum Hamiltonian, which, in turn, determines the ground-state wave function. The reverse of this ordering is also formally true, so let us start with a study of the lattice regularized form of the presumptive ground-state wave function, ${\widetilde\Psi}_0(\phi)\equiv\exp[-U(\phi)/2]$, where, as usual, $U(\p)$ is determined by the `large-field' behavior of the potential, and thus $U(\p)$ is well behaved when $\p$ is `small'. On the lattice, $\{\phi(x)\}$, at $t=0$, is replaced by $\{\p_k\}$, where $k\in{\mathbb Z}^s$ labels the site on a (spatial) hypercubic, periodic, lattice with lattice spacing $a>0$ and a total number of (spatial) sites $L^s\equiv N'<\infty$. The continuum limit arises when $a\ra0$, $L\ra\infty$, but $(L\s a)^s=N'\s a^s$ is fixed and finite, at least initially. Many moments of the ground-state distribution diverge in the continuum limit, such as
                  \bn \int [\Sigma'_k\phi_k^2\s]^p\,e^{\t-U(\phi)}\,\Pi'_k\s d\phi_k={\cal O}(N'^p)\;,\label{q1}\en
                  where the estimated value arises because there are $N'^p$ terms each of ${\cal O}(1)$, and as $N'\ra\infty$, divergences arise, for all $p\ge1$. [Remark: In (\ref{q1}) $\Sigma'_k$ and $\Pi'_k$ denote a sum and product over all sites in a single spatial slice.] These divergences seem to arise from the fact that the continuum limit involves an infinite number of integration variables, but the continuum limit need not lead to divergences. To understand this remark, let us first change the integration variables from ``Cartesian coordinates'' to ``hyperspherical coordinates'' by the transformation $\p_k\equiv\kappa\s\eta_k$, where $\kappa^2\equiv\Sigma'_k\p_k^2$, $1\equiv\Sigma'_k\s\eta_k^2$, $0\le\kappa<\infty$, and $-1\le\eta_k\le1$, for all $k$. In the new variables, (\ref{q1}) becomes
                    \bn \int \s[\k^2]^p\,e^{\t-U(\k\s\eta)}\,\k^{N'-1}\s d\k\,2\s\delta(1-\Sigma'_k\s\eta_k^2)\,\Pi'_k\s d\eta_k\;.\label{q2}\en
                  No longer do we have $N'^p$ terms of order ${\cal O}(1)$, but it is the power $N'-1$ of the hyperspherical radius $\k$ that leads to divergences as $N'\ra\infty$. Moreover,
                  a steepest descent analysis as $N'\ra \infty$---which makes the support of the measure in (\ref{q2})  {\it disjoint}, at least partially, for a change of parameters in $U$---leads to divergences in any perturbation analysis.
                  However, if the ground-state distribution contained an additional factor, namely, $\k^{-(N'-R)}$, where $R>0$ is fixed and finite, it  would effectively change the $\k$-factor from $\k^{N'-1}$ to $\k^{R-1}$---a procedure we call {\it measure mashing}---and, as a result, the divergences would disappear as $N'\ra\infty\s!$ Stated otherwise, measure mashing would nullify the steepest descent argument, and thus a perturbation analysis would not involve divergences.
                  In summary, the ``trick'' in securing a finite
                  version of scalar field quantization---one that also smoothly passes to its own pseudofree theory---is a direct result of mashing the measure.

                  To achieve measure mashing, we now assume the ground-state distribution has the form
                  \bn \Psi_0(\p)^2\equiv\{\s\Pi'_k [\Sigma'_l\s J_{k,l}\,\p_l^2\s]^{-(1-2ba^s)/2}\s\}\,e^{\t-U'(\p)}\;, \label{p1}\en
                  where $J_{k,l}\equiv 1/(2s+1)$ for $l=k$ and for $l$ equal to each site of the $2\s s$ spatially nearest neighbors of $k$; $J_{k,l}\equiv 0$ otherwise. Specifically, $R\equiv 2\s b\s a^s\s N'$, where $b>0$ has the dimensions of (length)$^{-s}$ to make $R$ dimensionless. In turn, the functional form of the lattice ground state determines the functional form of the lattice-regularized Hamiltonian operator. Roughly speaking, the denominator term fixes the small-field dependence of the potential, while $U'(\phi)$ leads to the
                  large-field behavior of the potential. However, to fix the Hamiltonian, we let the chosen denominator (involving $J_{k,l}$) of the ground-state wave function determine the small-field potential, but for the large-field behavior, we specify the form of the Hamiltonian  itself. This leads us to the lattice regularized form of the quantum Hamiltonian operator given by
                \bn &&\H=-\half\s a^{-2s}\s\hbar^2{\ts\sum}'_k\frac{\d^2}{\d\p_k^2}\s\s a^{s}+\half {\ts\sum}'_{k,k^*}\s(\p_{k^*}-\p_k)^2\s\s a^{s-2}+\half\s m_0^2{\ts\sum}'_k\s\p_k^2\s\s a^s\no\\
                &&\hskip3em+g_0{\ts\sum}'_k\s\p_k^4\s\s a^s+\half\s\hbar^2{\ts\sum}'_k{\cal F}_k(\p)\s\s a^s-E_0\;, \label{h1}\en
                where $k^*$ denotes each of the $s$ nearest neighbors to $k$ in the positive sense, and the all-important counterterm ${\cal F}_k(\p)$ is given by
               \bn {\cal F}_k(\p)\hskip-1,3em&&=\quarter\s(1-2ba^s)^2\s
          a^{-2s}\s\bigg({\ts\sum'_{\s t}}\s\frac{\t
  J_{t,\s k}\s \p_k}{\t[\Sigma'_m\s
  J_{t,\s m}\s\p_m^2]}\bigg)^2\no\\
  &&\hskip2em-\half\s(1-2ba^s)
  \s a^{-2s}\s{\ts\sum'_{\s t}}\s\frac{J_{t,\s k}}{[\Sigma'_m\s
  J_{t,\s m}\s\p^2_m]} \no\\
  &&\hskip2em+(1-2ba^s)
  \s a^{-2s}\s{\ts\sum'_{\s t}}\s\frac{J_{t,\s k}^2\s\p_k^2}{[\Sigma'_m\s
  J_{t,\s m}\s\p^2_m]^2}\;. \label{eF} \en
  Although ${\cal F}_k(\p)$ does not depend only on $\p_k$, it nevertheless becomes a local potential
  in the formal continuum limit. The constant $E_0$ is chosen to ensure that $\H\Psi_0(\p)=0$. Note that local field powers do not involve normal ordering, but instead, they are defined by an operator product expansion, realized effectively by a multiplicative renormalization of the parameters  \cite{JRK}.

At this point, the reader may be wondering what is the relation of enhanced quantization, the topic in the first part of this article, and the approach taken to describe scalar field quantization in the second part of this article. In the first part we introduced coherent states, both canonical and affine. Coherent states of any kind are normally based on a fiducial vector, i.e., our $|\eta\>$, and for a field theory with an infinite number of degrees of freedom, the usual fiducial vectors (e.g., Gaussian functions) are generally inappropriate. A safe fiducial vector to use is the ground state of the associated Hamiltonian, and for the scalar field case under discussion, the ground-state wave function, as given by the square root of (\ref{p1}), leads to affine coherent states being appropriate \cite{aff}, and thus our scalar field analysis is a form of affine quantization.

From the Hamiltonian operator we can determine the form of the (Euclidean) lattice action functional as given by
    \bn &&\hskip-2em I=\half \Sigma_{k,k^*}(\p_{k^*}-\p_k)^2\s\s a^{n-2}+\half\s m_0^2\Sigma_k\s\p_k^2\s\s a^n+g_0\s\Sigma_k\s\p^4_k\s\s a^n
    +\half\s\hbar^2\s\Sigma_k{\cal F}_k(\p) \s\s a^n \;, \label{t1}  \no\\ &&  \en
    where $n=s+1$, $\Sigma_k$ signifies a sum over the entire, finite, $n$-dimensional spacetime lattice, and now $k^*$ runs over all $n$ nearest neighbors to $k$ in the positive sense. The Euclidean-spacetime generating functional is given, in turn, by
      \bn &&S(h)=M\int e^{\t Z^{-1/2}\Sigma_k\s h_k\s\p_k\s a^n/\hbar-I/\hbar}\;\Pi_k\s d\p_k\;, \en
      where $Z$ denotes the field-strength renormalization constant, $\Pi_k$ is a product over all sites in the (finite) spacetime lattice, and $M$ is chosen so that $S(0)=1$. Based on the distribution underlying this integral, preliminary Monte Carlo studies \cite{JS} show a positive, non-vanishing renormalized coupling constant vs.~the bare coupling constant for $\p^4_4$. This result compares with an apparently  vanishing renormalized coupling constant vs.~the bare coupling constant that follows from a Monte Carlo study of conventional canonical quantization procedures \cite{FREE}.
 \subsection{Discussion of scalar field quantization}
      The counterterm ${\cal F}_k(\p)$ does not depend on $g_0$ and thus it remains as $g_0\ra0$. The result of that limit is the pseudofree model, and it differs from the usual free model. This kind of interacting theory provides a valid quantization of the non-renormalizable models such as $\p^4_n$, $n\ge5$. However, we can also extend the use of the new Hamiltonian operator to lower dimensions as well, even though, for the classical theory, the free and pseudofree models are the same. The purpose of extending the new form of the Hamiltonian is to ensure the uniqueness of `mixed models'. For example, consider the case of $\p^4_3$, which is a super-renormalizable model that has been studied perturbatively and non-perturbatively with the same self-consistent results; e.g., see \cite{GJ}. But, suppose we studied the mixed model given by $g_0\s\p^4_3+g'_0\s\p^8_3$, which is a sum of a super-renormalizable and a non-renormalizable model, or the mixed model $g''_0\s\p^4_5+g'''_0\s\p^8_5$, which is the sum of two different non-renormalizable models, etc. Conventional canonical quantization procedures would not be able to make any sense of such mixed models, but the new version, which treats each ingredient of such models in the same manner, would make perfectly good sense  as the coupling constants are turned on, then off, then on again, etc., in any order. In this regard we are also proposing new and different quantization procedures for models like $\p^4_2$ and $\p^4_3$, and other super-renormalizable models. [Remark: Of course, the original solutions for super-renormalizable models have their own role to play for different physical situations.] On these grounds, we advocate accepting the lattice Hamiltonian (\ref{h1}) and the lattice action (\ref{t1}) for all $n\ge2$.

\section*{Acknowledgements}
       Thanks are expressed to the Zentrum f\"ur interdisziplin\"are Forschung, Bielefeld University, for their support in attending the conference ``Stochastic and Infinite Dimensional Analysis''; and
       it is a pleasure to congratulate Prof.~Ludwig Streit on his 75th birthday.


\begin{thebibliography} {99}
      \bibitem{dirac}P.A.M.~Dirac, {\it The Principles of Quantum Mechanics}, (Clarendon Press,  Oxford, 1947), page 114.

   \bibitem{enh} J.R.~Klauder, ``Enhanced Quantization: A Primer'',  J. Phys. A: Math. Theor. {\bf 45}, 285304 (8pp) (2012); arXiv:1204.2870.

      \bibitem{skag-kla}J.R.~Klauder and B.-S.~Skagerstam, {\it Coherent States:  Applications to Physics and Mathematical Physics},  (World Scientific, Singapore, 1985).

      \bibitem{rs} J.R.~Klauder, ``Rotationally-Symmetric Model Field
                    Theories", J. Math. Phys. {\bf 6}, 1666-1679 (1965); H.D.I.~Abarbanel, J.R.~Klauder, and J.G.~Taylor, ``Green's Functions for Rotationally Symmetric
   Models", Phys.~Rev.~{\bf 152}, 198-206 (1966).

      \bibitem{GJ} J.~Glimm and A.~Jaffe, {\it Quantum Physics}, (Springer Verlag, New York, 1987), Second edition.

      \bibitem{JRK} J.R.~Klauder, ``Scalar Field Quantization Without Divergences In All Spacetime Dimensions'',
   J. Phys. A: Math. Theor. 44, 273001 (30pages) (2011); arXiv:1101.1706; J.R.~Klauder, ``Divergences in Scalar Quantum Field Theory: The Cause and the Cure'',
     Mod. Phys. Lett. A {\bf 27}, 1250117 (9pp) (2012); arXiv:1112.0803.


      \bibitem{lady}O.A.~Ladyzenskaja, V.~Solonnikov,  and N.N.~Ural'ceva, {\it Linear and
     Quasi-linear Equations of Parabolic Type}, (Am.~Math.~Soc., Providence, Vol.~23, 1968).

      \bibitem{book1}  J.R.~Klauder, {\it Beyond Conventional Quantization}, (Cambridge University Press,
Cambridge, 2000 \& 2005).

      \bibitem{aff}  J.R.~Klauder, ``New Affine Coherent States Based on Elements of Nonrenormalizable Scalar Field Models'', Advances in Mathematical Physics, vol. 2010, Article ID 191529, 16 pages, (2010);
http://www.hindawi.com/journals/amp/2010/191529.html;   J.R.~Klauder,``The Utility of Affine Variables and Affine Coherent States'',
     J. Phys. A: Math. Theor. {\bf 45}, 24001 (17pp) (2012);
         arXiv:1108.3380.

         \bibitem{JS} J.~Stankowicz, private communication.  

    \bibitem{FREE} B.~Freedman, P.~Smolensky, and
    D.~Weingarten, ``Monte Carlo Evaluation of the Continuum Limit of $\p^4_4$ and $\p^4_3$'',
    Phys. Lett. B {\bf 113}, $481-486$ (1982).

    %
    %
     %
\end{thebibliography}
\end{document}